\documentclass[aps,preprint,prb,groupedaddress,showpacs,graphics]{revtex4}
\usepackage{graphicx}
\usepackage{amsmath}

\newcommand{\bvec}{\mathbf b}

\newcommand{\evec}{\mathbf e}
\newcommand{\kvec}{\mathbf k}
\newcommand{\rvec}{\mathbf r}
\newcommand{\qvec}{\mathbf q}

\newcommand{\Avec}{\mathbf A}
\newcommand{\Bvec}{\mathbf B}
\newcommand{\Evec}{\mathbf E}
\newcommand{\Fvec}{\mathbf F}

\newcommand{\dkvec}{\text{d}\mathbf k}
\newcommand{\drvec}{\text{d}\rvec}
\newcommand{\dqvec}{\text{d}\qvec}
\newcommand{\domega}{\text{d}\omega}

\newcommand{\dz}{\text{d}z}
\newcommand{\ket}[1]{\mid #1 \rangle}

\newcommand{\braket}[2]{\langle #1 \mid #2 \rangle}

\newcommand{\Acal}{\mathcal{A}}

\newcommand{\Fcal}{\mathcal{F}}

\newcommand{\Pcal}{\mathcal{P}}

\newcommand{\Vcal}{\mathcal{V}}

\newcommand{\DfDx}[2][{}]{\frac{\partial #1}{\partial #2}}
\newcommand{\DDfDxx}[2][{}]{\frac{\partial^2 #1}{\partial #2^2}}
\newcommand{\Tblip}{\ensuremath{T_{\textrm{blip}}}}

\begin{document}

\title{Sequential multi-photon strategy for semiconductor-based terahertz detectors}

\author{Fabrizio Castellano, Rita C. Iotti, and Fausto Rossi}
\affiliation{Dipartimento di Fisica, Politecnico di Torino, Corso
Duca degli Abruzzi 24, 10129 Torino, Italy}

\begin{abstract}
A semiconductor-based terahertz-detector strategy, exploiting a
bound-to-bound-to-continuum architecture, is presented and
investigated. In particular, a ladder of equidistant energy levels
is employed, whose step is tuned to the desired detection
frequency and allows for sequential multi-photon absorption. Our
theoretical analysis demonstrates that the proposed multi-subband
scheme could represent a promising alternative to conventional
quantum-well infrared photodetectors in the terahertz spectral
region.
\end{abstract}

\maketitle

\section{Introduction}

The recent development of reliable far-infrared (far-IR)
semiconductor-based laser sources, such as the quantum-cascade
(QC) laser\cite{Nature,Rochat}, together with the potential
applications in imaging, communication and medicine, identify
terahertz (THz) radiation detection as a crucial technological
milestone. To this end, many approaches have been proposed in the
last years, which aim at accessing the 1-10 THz region of the
electromagnetic spectrum. Currently proposed solutions encompass a
variety of different approaches, each with its own peculiar
characteristics.

From the electronics world, field effect transistors are extending
their operation frequency into the sub-THz and THz region
exploiting plasmon resonance effects\cite{Knap1,Knap2,Shaner}. On
the other hand, optoelectronic techniques benefitting from
electro-optical properties of LiTaO$_3$, LiNbO$_3$ and ZnTe
crystals have been proposed\cite{Winnewisser,Nahata}.

Semiconductor heterostructures also play a fundamental role in
this field, QC structures\cite{Hof,Graf} as well as quantum-well
infrared photodetectors (QWIPs)\cite{QWIP} being among the most
promising directions. Concerning the latter, radiation detection
via conventional QWIP designs resorts on direct bound-to-continuum
electronic transitions, which allowed to achieve remarkable levels
of performance in the mid-IR range. Recently, the use of
multi-level architectures, opening up to bound-to-bound electronic
transitions, has been proposed and studied, focusing both on their
intrinsic nonlinear character and on their wide-band absorption
spectra. While the latter feature allows for
multi-color\cite{multi2} or wideband
detection\cite{multi5,multi6,multi8,multi9}, second-order
nonlinearities of two-level systems have been studied and
experimentally demonstrated with the idea of using the devices for
second-order autocorrelation
measurements.\cite{multi1,multi3,multi4,multi7}.

The extension of the conventional, bound-to-continuum, QWIP
principle into the far-IR range is not straightforward. In
particular, one of the main issues in THz-operating devices are
the huge dark current values that cause the background limited
infrared photodetection temperature (\Tblip) to be in the range
10--15 K,\cite{Liu, Liu2} that is, much lower than that of
state-of-the-art mid-IR QWIPs. In a previous work,\cite{multi10}
we addressed the advantages of the application of multi-level
architecture in THz QWIP designs, and concluded that a
bound-to-bound-to-continuum scheme may efficiently face the
above-mentioned dark current issue. More recently\cite{multi11},
we have analyzed the performances of such novel architecture,
focusing on the characteristic figure of merit \Tblip. Our results
suggest the possibility to achieve a consistent improvement of the
operation temperature of THz QWIPs by means of our proposed
multi-level design. In the present article, our findings are
further discussed and the theoretical model on which our
calculations are based is explained in more detail.

\section{Physical systems and modeling strategy}
Our prototypical device consists of an infinitely periodic
semiconductor-based heterostructure supporting, within each
period, a set of equally-spaced bound states.
%
%
The physical system we are considering is therefore an electron
gas within a periodic nanostructure and in the presence of
external electromagnetic fields. The corresponding Hamiltonian can
be schematically written as
\begin{equation}
\label{hamiltonian} \hat{H} =  \hat{H}^\circ + \hat{H}^\prime \; .
\end{equation}
The first term of Eq.~(\ref{hamiltonian}),
\begin{equation}
\label{wf}
 \hat{H}^\circ = \hat{H}^\circ_e + \hat{H}^\circ_{qp} = \sum_\alpha \epsilon_\alpha \hat{c}^\dagger_\alpha \hat{c}_\alpha + \sum_{\lambda\qvec} \epsilon_{\lambda\qvec} \hat{b}^\dagger_{\lambda\qvec} \hat{b}_{\lambda\qvec}
\end{equation}
is the sum of the free-carrier ($\hat{H}^\circ_e$) and
free-quasiparticle ($\hat{H}^\circ_{qp}$) Hamiltonians, where the
fermionic operator $\hat{c}^\dagger_\alpha$ ($\hat{c}_\alpha$)
denotes creation (destruction) of a carrier in the single particle
state $\alpha$, with energy $\epsilon_\alpha$, while the bosonic
operator $\hat{b}^\dagger_{\lambda\qvec}$
($\hat{b}_{\lambda\qvec}$) denotes creation (destruction) of a
quasiparticle excitation of type $\lambda$ (phonons, photons,
plasmons, etc) with wave vector $\qvec$.

The Hamiltonian $\hat{H}^\prime$ in Eq.~(\ref{hamiltonian}) is the
sum of all possible interaction terms between electrons and
quasiparticles. Since the aim of the present paper is to provide a
focus on the electron-photon interaction dynamics, the latter will
be treated in a fully microscopic scheme, in terms of the Fermi's
golden rule. Conversely, all the other carrier-quasiparticle
interactions will be described within a phenomenological
electronic mean-lifetime picture, providing effective scattering
probabilites that guarantee the proper thermalization of the
electron population in the absence of external electromagnetic fields.

\subsection{Band structure calculation}
The single-particle Hamiltonian $\hat{H}_e^\circ$ describes the
non-interacting carrier system within the effective
three-dimensional potential profile of our quantum device. The
generic label $\alpha$ adopted in Eq.~(\ref{wf}) denotes, in
general, a suitable set of discrete and/or continuous quantum
numbers; for the case of quasi-two-dimensional semiconductor
heterostructures, as the ones considered in this paper, the latter
includes a partially discrete index along the so-called growth
direction. In particular, since our prototypical design is made up
of a sequence of identical units, the potential term consists, in
the envelope-function formalism, of a periodic one-dimensional
(1D) profile.

For a device grown along the $z$-direction and homogeneous as far
as the in-plane ($x,y$) dynamics is concerned, the following
factorization of the electron wavefunction may then be assumed
\begin{equation}
\label{psifact} \Psi_{bk_z{\kvec_p}}(\rvec) =
\psi_{b,k_z}(z)\phi_{\kvec_p}(x,y) \; ,
\end{equation}
where $\kvec_p$ and $k_z$ are the in-plane and along-$z$
components of the electron wavevector $\kvec$, respectively, and
$b$ is the label numbering the various discrete subbands in which
the conduction band is split because of the 1D quantum confinement
potential.

While parabolic bands are considered for the in-plane dispersion,
and $\phi_{\kvec_p}(x,y)$ is the corresponding plane wave, the
band structure along the growth direction is computed from the 1D
Schr\"odinger equation for the given potential profile. Moreover,
due to the typically low doping levels in this kind of devices,
charge-density effects on the potential profile may safely be
neglected and no Schr\"odinger-Poisson coupling is included in our
modeling.

The Schr\"odinger equation projected along the $z$ direction is
solved by means of a plane-wave expansion, as described in
Ref.~[\onlinecite{BBR}]. The following basis functions may then be
adopted
\begin{equation}
\label{basis} \chi_{n,k_z}(z) = \frac{1}{\sqrt{L_z}}e^{i(G_n +
k_z)z}
\end{equation}
where $n$ is an integer running from $-N$ to $N$, $L_z$ is the
period of the 1D potential (i.e., the supercell width), and $G_n =
2\pi n/L_z$ and $k_z$ ($-\pi/L_z < k_z < \pi/L_z$) are the
reciprocal lattice vector and the quasimomentum in the first
Brillouin zone, respectively. The basis functions are
normalized, as usual, over the supercell
\begin{equation}
 {\int_{-L_z/2}^{L_z/2}\chi^*_{n,k_z}\chi_{m,k_z}\dz} = \delta_{nm} \; .
\end{equation}

In a reduced-zone scheme we can express the along-$z$ wave
function of an electron in subband $b$ and momentum $k_z$ as
\begin{equation}
\label{psi} \psi_{b,k_z}(z) =
\sqrt{\frac{L_z}{2\pi}}\sum_{n=-N}^{N} c_{b,n,k_z}\chi_{n,k_z}(z)
\; .
\end{equation}
The series expansion in Eq.~(\ref{psi}) allows us to convert the
stationary Schr\"odinger equation into a discrete eigenvalue
problem. The solution of such a problem consists of a set of
$2N+1$ energy eigenvalues, $\epsilon_{b,k_z}$, each representing the
allowed energy level for an electron in subband $b$ with wavevector $k_z$.
The components $c_{b,n,k_z}$ represent the
spectrum of the wavefunction in the plane wave basis set. The
plane-wave like normalization of the wavefunctions
$\Psi_{b,k_z,\kvec_p}$,
\begin{equation}
 \langle \Psi_{b k_z'\kvec_p'} \vert \Psi_{b k_z\kvec_p}\rangle =
  \int \Psi^*_{b k_z'\kvec_p'}({\bf r})\Psi_{b k_z\kvec_p}({\bf r}) \, d{\bf r}
  = \delta(\kvec - \kvec') \; ,
\end{equation}
is guaranteed by the form~(\ref{basis}) of the basis functions and
is consistent with the fact that the structure is assumed to be
infinite along $z$.

\subsection{Potential profile}
The quantum design of our semiconductor device should satisfy
several requirements.
First of all, the main constrain is to have equally spaced bound levels.
Secondly, we want to be able to control the number of such levels and
their spacing too.

When speaking of equally spaced levels, the first solution would
seem to be that of a parabolic potential profile. The
implementation of the latter, however, besides non-trivial growth
issues, poses more fundamental problems: in order to have carrier
transport we need a continuum and thus the parabolic potential
must be truncated at some point. Such a truncated parabola would
not support equally spaced levels anymore. We therefore decide to
use multi-quantum-well strategies for our QWIP basic period.

Single quantum wells are used to produce the single bound level
providing the bound-to-continuum transition exploited in
conventional QWIPs. Two energetically equal transitions can still
be obtained with a single QW of proper geometry. The tuning of the
separations of three bound levels cannot be achieved with a
potential having only two free parameters (width and depth) and
thus we have to switch to more complex structures.

\begin{figure}
  \includegraphics*[width=\linewidth,height=.8\linewidth]{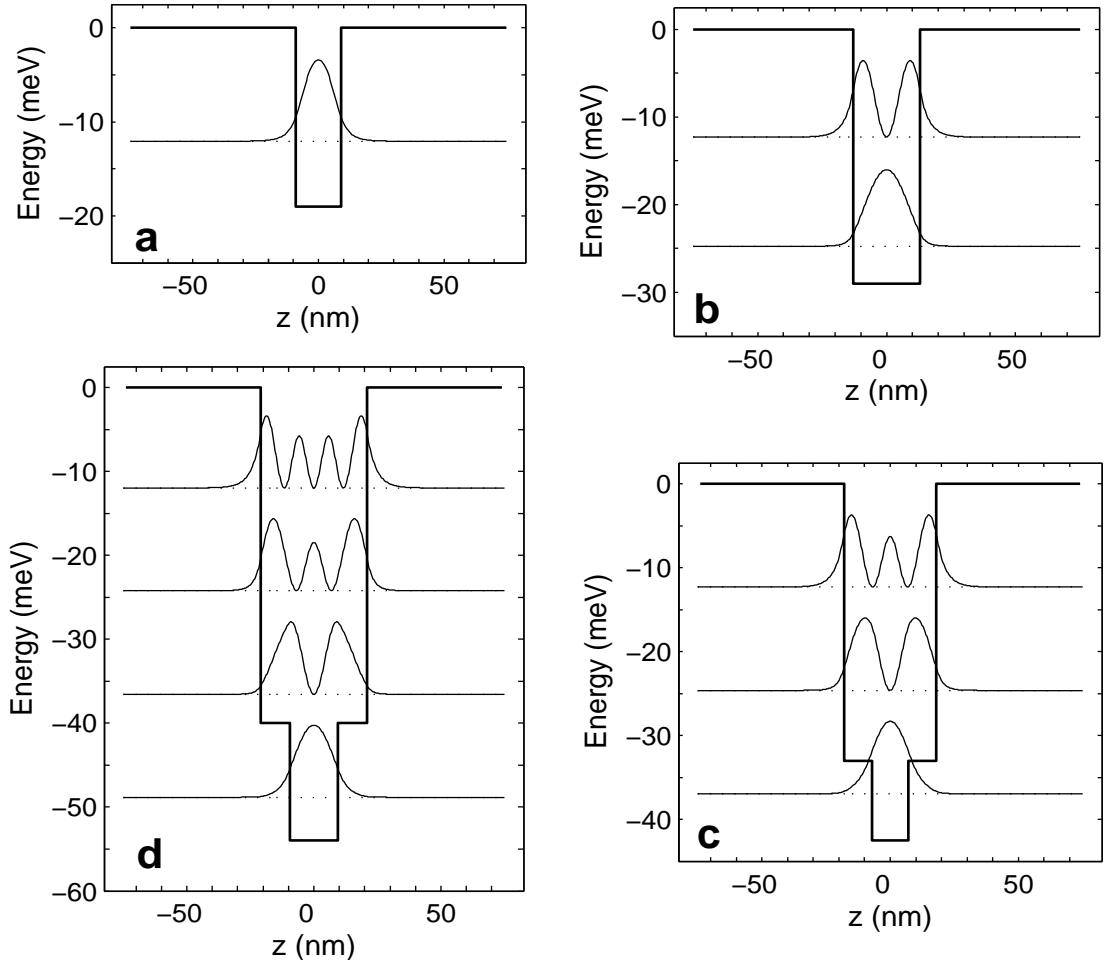}
  \caption{Potential profiles along the growth direction of our prototypical devices, designed to operate at 3 THz,
  with a number of bound states varying from one (a) to four (d).
The proposed symmetric nested-quantum-well structure, in (c) and
(d), provides additional geometric parameters, with respect to the
single QW design, that can be varied to tune the energy level
separation.}\label{profile}
\end{figure}

The nested QW structures ---shown in Figure~\ref{profile}-- turn
out to be convenient choices. The introduction of additional
geometrical parameters to the standard QW design allow us to
control the number and position of the desired number of energy
levels. A detailed description of the method used to determine the
potential profiles is given in Appendix~\ref{appA}.

Fig.~\ref{profile} shows the supercells of our prototypical structures
which are to be infinitely
replicated along the growth direction. The use of many repetitions
of the basic unit is indeed the strategy exploited in this kind of
unipolar devices to optimize detection efficiency. Finite-size
(i.e., boundary and contact) effects are therefore of minor
importance. Moreover, due to the low doping values, in-plane
quantum confinement effects are negligible too.

\subsection{Transport model}
The transport model we employ, to describe the electron dynamics
in our unipolar device, is based on the Boltzmann transport
equation describing the distribution of electrons in the device
conduction band. Its general form for the case of $N$ subbands is
the following
\begin{widetext}
\begin{equation}
\label{boltzmann1}
\DfDx[f_b(\kvec)]{t} = \frac{e}{\hbar}\Fvec\cdot\nabla
f_b(\kvec) + \sum_{b' =
1}^N\int\left[P_{bb'}(\kvec,\kvec')f_{b'}(\kvec') -
P_{b'b}(\kvec',\kvec)f_b(\kvec)\right]\dkvec'
\end{equation}
\end{widetext}
where $f_b(\kvec)$ is the single-particle distribution function of
electrons in a state with wavevector $\kvec$ in subband $b$,
$P_{b'b}(\kvec',\kvec)\dkvec'$ is the probability per unit time
that a scattering event bringing an electron from a state in band
$b$ and wavevector $\kvec$ to a state in band $b'$ and wavevector
$\kvec'$ occurs, and $\Fvec$ is the external electric field
providing the electron drift. $\Fvec$ may in general be oriented
in any direction; in this paper, we will limit our discussion to
biases applied only along the growth axis.

The knowledge of $f_b(\kvec)$ allows us to evaluate the current
density across the device, ${\cal J}$, as follows
\begin{equation}
\langle {\cal J}\rangle = \frac{e}{(2\pi)^3 \hbar}\sum_b\int
\nabla E_b(\kvec)f_b(\kvec)\dkvec \; ,
\end{equation}
provided that the distribution function is normalized as
\begin{equation}
\label{fnorm} \frac{1}{(2 \pi)^3}\sum_b\int f_b(\kvec)\dkvec = N_e
\; ,
\end{equation}
where $E_b(\kvec)$ is the miniband dispersion and $N_e$ the number
of electrons per unit volume in the device.

Being interested in the steady-state behavior of our device, we
solve the homogeneous equation obtained from Eq.~(\ref{boltzmann1})
when the time derivative is set equal to zero
\begin{widetext}
\begin{equation}
 \label{boltzmann2}
\frac{e}{\hbar}\Fvec\cdot\nabla
f_b(\kvec) + \sum_{b' =
1}^N\int\left[P_{bb'}(\kvec,\kvec')f_{b'}(\kvec') -
P_{b'b}(\kvec',\kvec)f_b(\kvec)\right]\dkvec'=0 \; .
\end{equation}
\end{widetext}
The latter equation is solved employing a finite difference
strategy as described in Appendix~\ref{appB}.

The various scattering mechanisms affecting the electron dynamics
are included into the global probabilities $P_{b'b}(\kvec',\kvec)$
and may be separated into the following contributions
\begin{equation}
\label{P} P_{b'b}(\kvec',\kvec) = P_{b'b}^{opt}(\kvec',\kvec) +
P_{b'b}^{th}(\kvec',\kvec) \; ,
\end{equation}
where $P_{b'b}^{opt}(\kvec',\kvec)$ is the electron-photon
interaction part and $P_{b'b}^{th}(\kvec',\kvec)$ accounts
for all thermalization processes.

\subsection{Non-optical scattering model}
\label{Nopt} To keep the model as simple as possible, yet without
spoiling the proper description of the main physical issues, all
non optical scattering processes are accounted for by means of a
phenomenological mean lifetime $\tau$ which acts as a global
fitting parameter.

Let us introduce a thermal transition probability density
$P^{th}_{bb'}(\kvec,\kvec')$ such that the mean lifetime
$\tau_{b\kvec}$ of an electron in band $b$ with wavevector $\kvec$
is given by
\begin{equation}
 \frac{1}{\tau_{b\kvec}} =  \sum_{b'}\int
 P^{th}_{b'b}(\kvec',\kvec)\dkvec' \; .
\end{equation}

The mean lifetime of the electrons, $\tau$, is then defined in
terms of the distribution function $f_b(\kvec)$ as
\begin{equation}
\label{tau1}
 \frac{1}{\tau} = \frac{1}{(2\pi)^3N_e} \sum_b\int \frac{f_b(\kvec)}{\tau_{b\kvec}}\dkvec
 = \frac{1}{(2\pi)^3N_e}
 \sum_{bb'}\iint f_b(\kvec)P^{th}_{b'b}(\kvec',\kvec)\, \dkvec'\dkvec
 \; .
\end{equation}
The latter can be used to compute the $P^{th}$ probabilities once
$\tau$ has been fixed and a functional form for $P^{th}$ has been
set. However, the definition of $\tau$ given in Eq.~(\ref{tau1})
implies the knowledge of the single-particle distribution
function, which is obtained from Eq.~(\ref{boltzmann2}), which in
turn requires $P^{th}$ to be determined. To break this loop we
choose to drop the strict physical interpretation of $\tau$ as the
actual mean lifetime of electrons and simply use it as a measure
of the strength of thermalization mechanisms. In this picture we
can perform the mean in (\ref{tau1}) using a distribution function
of our choice and convenience, bearing in mind that this won't
affect our conclusions. We thus define $\tau$ as
\begin{equation}
\label{tau2} \frac{1}{\tau} = \frac{1}{(2\pi)^3N_e}
\sum_{bb'}\iint P^{th}_{b'b}(\kvec',\kvec)\, \dkvec'\dkvec \; .
\end{equation}

To evaluate the integral in Eq.~(\ref{tau2}), we have to choose a
functional form for $P^{th}$ containing a free parameter suitable
for normalization. Since $P^{th}$ must account for all
thermalization mechanisms, its form must ensure that in the absence of
any external excitation (i.e., no bias, no light) the system
exhibits a thermal distribution function, that is, a distribution
function such that
\begin{equation}
\frac{f_b(\kvec)}{f_{b'}(\kvec')} = e^{-\frac{E_b(\kvec) -
E_{b'}(\kvec')}{k_BT}} \; .
\end{equation}
At equilibrium we know from the detailed-balance principle that
\begin{equation}
\frac{f_b(\kvec)}{f_{b'}(\kvec')} =
\frac{P^{th}_{bb'}(\kvec,\kvec')}{P^{th}_{b'b}(\kvec',\kvec)} \; ;
\end{equation}
the simpler way to fulfill this requirement is to impose
\begin{eqnarray}
\label{Pth}
 P^{th}_{b'\kvec',b\kvec} & = & P_0\Pcal_{b'b}(\kvec' \kvec) = \\
 \nonumber
& = &  P_0
\begin{cases}
\; 1 \qquad \qquad \quad\quad \text{ if } E_{b}(\kvec) > E_{b'}(\kvec') \\
  e^{-\frac{E_{b'}(\kvec') - E_{b}(\kvec)}{k_BT}} \quad \text{ if } E_{b}(\kvec) < E_{b'}(\kvec')
\end{cases}
\end{eqnarray}
where $P_0$ is a normalization constant that can be computed in
terms of $\tau$ as follows
\begin{equation}
 \frac{1}{P_0} = \frac{\tau}{(2\pi)^3N_e}\sum_{bb'}\int
 \Pcal_{b'b}(\kvec',\kvec)\dkvec'\dkvec \; .
\end{equation}

The strategy is therefore to first assume a value for $\tau$ and
then use the latter to compute $P_0$. This completely determines
the probabilities $P^{th}$ that appear in Eq.~(\ref{P}) and allows
us to solve Eq.~(\ref{boltzmann2}).

Actually the definition of $P_0$ would not be of any importance if
thermal scattering were the only scattering process, but, since we
want to investigate its competition/interplay with carrier-photon
interaction, $P_0$ (and consequently $\tau$) is the parameter that
allows us to adjust the relative strength of the two mechanisms.

\subsection{Electron-photon interaction}
To evaluate the \Tblip\ of our prototypical device, we have to
properly describe the interaction between the electron population
and the radiation field of an external blackbody source.

The second-quantization electric- and magnetic-field operators for a plane
electromagnetic wave with wavevector $\qvec$ have the form
\begin{gather}
\hat{\Evec}_\qvec = \frac{ \vert \Evec_\qvec \vert}{\sqrt{2}}
\evec_\qvec(e^{i(\omega_\qvec t - \qvec\cdot\rvec)}\hat{a}_\qvec +
e^{-i(\omega_\qvec t - \qvec\cdot\rvec)}\hat{a}^\dagger_\qvec) \\
\hat{\Bvec}_\qvec = \frac{ \vert \Bvec_\qvec \vert}{\sqrt{2}}
\bvec_\qvec(e^{i(\omega_\qvec t - \qvec\cdot\rvec)}\hat{a}_\qvec +
e^{-i(\omega_\qvec t - \qvec\cdot\rvec)}\hat{a}^\dagger_\qvec)
\end{gather}
or alternatively
\begin{gather}
\hat{\Evec}_\qvec = \sqrt{\frac{\hbar\omega_\qvec}{2\epsilon
\Vcal}}\evec_\qvec(e^{i(\omega_\qvec t - \qvec\cdot\rvec)}\hat{a}_\qvec
+ e^{-i(\omega_\qvec t - \qvec\cdot\rvec)}\hat{a}^\dagger_\qvec) \\
\hat{\Bvec}_\qvec = \sqrt{\frac{\hbar\omega_\qvec\mu}{2
\Vcal}}\bvec_\qvec(e^{i(\omega_\qvec t - \qvec\cdot\rvec)}\hat{a}_\qvec
+ e^{-i(\omega_\qvec t - \qvec\cdot\rvec)}\hat{a}^\dagger_\qvec)
\end{gather}
where $\Evec_\qvec$ and $\Bvec_\qvec$ are the classical electric
and magnetic fields, $\omega_\qvec$ is the dispersion relation of
the medium, $\varepsilon$ is the dielectric constant, $\mu$ the
magnetic permittivity, $\Vcal$ is the device volume, $\evec_\qvec$
and $\bvec_\qvec$ are the polarization unit vectors such that
$\evec_\qvec\cdot\qvec = \bvec_\qvec\cdot\qvec =
\evec_\qvec\cdot\bvec_\qvec = 0$, and  $\hat{a}_\qvec$ and
$\hat{a}^\dagger_\qvec$ are destruction and creation operators,
respectively, for a photon of wavevector $\qvec$.

The expressions above allow us to write the electric and magnetic
field operators in the case of a linear superposition of plane
waves as
\begin{gather}
 \hat{\Evec} = \sum_\qvec \hat{\Evec}_\qvec \\
\hat{\Bvec} = \sum_\qvec \hat{\Bvec}_\qvec \; .
\end{gather}
With the latter definition, we can easily recover the usual
expression for the second quantization hamiltonian of a population
of photons in terms of the energy density operator $\hat{U}(\rvec)$
\begin{equation}
\begin{split}
 \hat{H}^\circ_{ph} &= \int_\Vcal \hat{U}(\rvec)\drvec =\\
 &= \int_\Vcal \sum_\qvec
 \left( \frac{1}{2}\varepsilon \hat{\Evec}_\qvec\cdot \hat{\Evec}_\qvec^\dagger
+ \frac{1}{2\mu} \hat{\Bvec}_\qvec\cdot \hat{\Bvec}_\qvec^\dagger
 -\frac{\hbar\omega_\qvec}{2\Vcal} \right)\drvec = \\
 &= \sum_\qvec\hbar\omega_\qvec \hat{a}^\dagger_\qvec \hat{a}_\qvec \; .
\end{split}
\end{equation}
In this picture, the classical energy density, $U(\rvec) =
\frac{1}{2}\varepsilon \sum_\qvec E_\qvec^2 + \frac{1}{2\mu}
\sum_\qvec B_\qvec^2$, refers to the zero-point energy density,
$\frac{\hbar\omega_\qvec}{2\Vcal}$, of the electromagnetic field
in a cavity of volume $\Vcal$.

Given the electric field operator $\hat{\Evec}$, we can define the
vector potential operator $\hat{\Avec}$ as
\begin{equation}
 \hat{\Avec} = \sum_\qvec \hat{\Avec}_\qvec = \sum_\qvec \frac{1}{\sqrt{2}}(\Avec_\qvec \hat{a}_\qvec + \Avec_\qvec^*\hat{a}^\dagger_\qvec) = \sum_\qvec
 \frac{\hat{\Evec}_\qvec}{i\omega_\qvec} \; ,
\end{equation}
where $\Avec_\qvec$ is the classical vector potential, having
implicitly assumed a gauge where $\Evec = \DfDx[\Avec]{t}$.

In a second-quantization picture, the electron-photon interaction hamiltonian operator
\begin{equation}
 \hat{H}^{opt} = -i\hbar\frac{e}{m}\nabla\cdot \hat{\Avec} =
 -\frac{i\hbar e}{m\sqrt{2}}\sum_\qvec(\nabla\cdot\Avec_\qvec \hat{a}_\qvec + \nabla\cdot\Avec^*_\qvec \hat{a}^\dagger_\qvec)
\end{equation}
can be written as
\begin{equation}
\label{Hi}
 \hat{H}^{opt} = \sum_{\alpha\alpha'\qvec}\left[g_{\alpha\alpha'\qvec}\hat{c}^\dagger_\alpha \hat{a}_\qvec \hat{c}_{\alpha'} +
 g^*_{\alpha\alpha'\qvec}\hat{c}^\dagger_{\alpha'} \hat{a}^\dagger_\qvec \hat{c}_\alpha \right]
\end{equation}
here, the first (second) contribution describes a process in which
an electron performs a transition between the two single-particle
states $\alpha = (b,k_z,\kvec_p)$ and $\alpha'=(b',k_z',\kvec_p')$
absorbing (emitting) a photon; this mechanism has a coupling
constant $g$ which is expressed as
\begin{equation} \label{g}
 g_{\alpha\alpha'\qvec} = -\frac{i\hbar
 e}{m\sqrt{2}}\int\drvec\Psi_\alpha^*(\nabla\cdot\Avec_\qvec)\Psi_{\alpha'}
 \; .
\end{equation}

The evaluation of $ g_{\alpha\alpha'\qvec}$ from Eq.~(\ref{g}) can
be carried out in terms of the plane wave expansion of
$\Psi_\alpha$ given in Eq.s~(\ref{psifact}) and (\ref{psi})
\begin{widetext}
\begin{equation}
\label{g1}
 g_{bk_z \kvec_p,b'k_z' \kvec_p',\qvec} = \frac{\hbar e}{m\sqrt{2}} \, \delta(\kvec + \qvec - \kvec')
\sum_nc^*_{b'nk_z'}c_{bnk_z}\left[A_{z,\qvec}(q_z + G_n + k_z) +
\Avec_{p,\qvec}\cdot\kvec_p\right] \; ,
\end{equation}
\end{widetext}
where $A_{z,\qvec}$ and $\Avec_{p,\qvec}$ are the along-$z$
and in-plane components of the vector potential, respectively.

Equation~(\ref{g1}) may be simplified in several ways. First of
all, the usual dipole approximation allows to neglect the photon
momentum $\qvec$ with respect to the electron momentum $\kvec$.

Since we have assumed a parabolic in-plane dispersion, and since
we expect electrons to have a quasi-thermal distribution, then the
great majority of them will occupy states close to the subband
bottom ($k_p \approx 0$). On the other hand, the minibands
along $k_z$ are either flat or slightly dispersive, that is, much
narrower that the related subbands. Therefore we may assume that,
for the majority of the electrons, $k_p \ll (G_n + k_z)$.

This leads to the following simplified expression for the coupling
constant
\begin{equation}
\label{g2}
 g_{\alpha\alpha',\qvec} = \frac{\hbar e}{m\sqrt{2}}\delta(\kvec_\alpha - \kvec_{\alpha'})|\Avec_\qvec|\cos\varphi_\qvec
\sum_nc^*_{b'nk_z'}c_{bnk_z}(G_n + k_z)
\end{equation}
where $\varphi_\qvec$ is the angle between the vector potential
and the $z$ direction. The relevant term in the computation of
transition probabilities is $|g_{\alpha\alpha',\qvec}|^2$ which
contains a $\cos^2\varphi_\qvec$ term. If we consider a blackbody
radiation we can assume it as composed of a superposition of plane
waves with random polarization and thus we would replace
$\cos^2\varphi_\qvec$ with its mean value over $(0,2\pi)$, that is
$1/2$. Anyway each electromagnetic mode is the sum of two
independent polarizations thus we may simply replace
$\cos^2\varphi_\qvec \approx 1$, obtaining
\begin{equation}
\label{g3}
 g_{\alpha\alpha',\qvec} =
 \frac{\hbar e}{m\sqrt{2}}\delta(\kvec_\alpha - \kvec_{\alpha'})|\Avec_\qvec|p_{\alpha\alpha'}
\end{equation}
where
\begin{equation}
 p_{\alpha\alpha'} = \sum_nc^*_{b'nk_z'}c_{bnk_z}(G_n + k_z)
\end{equation}
is the matrix element of the momentum operator between states
$\alpha'$ and $\alpha$.

Let us now consider a photon absorption process, bringing the
system from state $\ket{\alpha',n_\qvec}$, with an electron in
state $\alpha'$ and $n$ photons with wavevector $\qvec$, to state
$\ket{\alpha, n_\qvec-1}$, with the electron in state $\alpha$ and
$(n-1)$ photons in state $\qvec$. Its probability per unit time
can be evaluated by Fermi's golden rule as
\begin{equation}
 P^{opt}_{\alpha\alpha',\qvec} =
 \frac{2\pi}{\hbar}|\braket{\alpha,n_{\bf q}-1}
 {\hat{H}^{opt}|\alpha',n_{\bf q}}|^2\delta(E_\alpha - E_{\alpha'} -
 \hbar\omega_\qvec) \; .
\end{equation}
The calculation gives
\begin{equation}
\label{Paaq1}
 P^{opt}_{\alpha\alpha',\qvec} =
 \frac{2\pi}{\hbar}|g_{\alpha\alpha',\qvec}|^2n_\qvec\delta(E_\alpha - E_{\alpha'} -
 \hbar\omega_\qvec) \; .
\end{equation}
On the other hand, the probability of a photon emission process,
in which the system performs a transition from state
$\ket{\alpha,n_\qvec}$ to state $\ket{\alpha', n_\qvec +1}$, is
\begin{equation}
\label{Paaq2}
 P^{opt}_{\alpha'\alpha,\qvec} =
 \frac{2\pi}{\hbar}|g_{\alpha'\alpha,\qvec}|^2(n_\qvec + 1)\delta(E_\alpha - E_{\alpha'} +
 \hbar\omega_\qvec) \; .
\end{equation}

\subsection{Interaction with blackbody radiation}
Since our aim is to determine the \Tblip\ of our prototypical
detector, we need to study its interaction with the background
radiation, considered as a blackbody radiation at 300 K. From a
quantum mechanical point of view, a blackbody radiation is a
photon population at thermal equilibrium following the
Bose-Einstein distribution law.

A non interacting electron system only coupled to a photon bath at
thermal equilibrium, must itself thermalize. Indeed, by employing
the detailed-balance principle and substituting the Bose-Einstein
distribution in Eqns.~(\ref{Paaq1}) and~(\ref{Paaq2}) we can write
\begin{equation}
 \frac{f_\alpha}{f_{\alpha'}}
 = \frac{P^{opt}_{\alpha\alpha',\qvec}}{P^{opt}_{\alpha'\alpha,\qvec}}
 = \frac{n_\qvec}{(n_\qvec + 1)} = e^{-\frac{\hbar\omega_\qvec}{k_B T}} = e^{-\frac{E_\alpha -
 E_{\alpha'}}{k_B T}} \; ,
\end{equation}
that is, the steady-state distribution function is such that the
ratio between the occupation numbers of states $\alpha$ and
$\alpha'$ is, as expected, the Boltzmann factor.

Equations (\ref{Paaq1}) and (\ref{Paaq2}) give the transition
probabilities for an electron interacting with an electromagnetic
plane wave, which can be seen as an electromagnetic mode of a
cavity. When our device is inside a cavity at thermal equilibrium
(a blackbody), the total transition probabilities must be summed
over all modes $\qvec$. This is also formally described by the
interaction hamiltonian (\ref{Hi}) which is a sum over all
wavevectors $\qvec$. We therefore write, for the absorption
process,
\begin{equation}
 P^{opt}_{\alpha\alpha'} =
 \sum_\qvec P^{opt}_{\alpha\alpha',\qvec} =
 \frac{2\pi}{\hbar} \sum_\qvec |g_{\alpha\alpha',\qvec}|^2n_\qvec\delta(E_\alpha - E_{\alpha'} -
 \hbar\omega_\qvec) \; .
\end{equation}
In the limit of a infinitely large cavity, the summation becomes
an integral in $\dqvec$ and $|g_{\alpha'\alpha,\qvec}|^2$ becomes
a spectral density $|g_{\alpha'\alpha}(\qvec)|^2$ which is related
to the squared vector potential spectral density
$|\Acal(\qvec)|^2$ through equation~(\ref{g3}).

The quantity $|\Avec(\qvec)|^2$ can be expressed in terms of the
energy density $U(\qvec) =
\frac{1}{2}\varepsilon|\Evec(\qvec)|^2 + \frac{1}{2\mu}|\Bvec(\qvec)|^2$ and, considering the
relations $\Evec(\qvec) = i\omega_\qvec \Avec(\qvec)$, $\Bvec = \nabla\times\Avec$, as
\begin{equation}
\label{A}
 |\Avec(\qvec)|^2 = \frac{U(\qvec)}{\varepsilon
 \omega_\qvec^2} \, .
\end{equation}
For a linear dispersion relation
$\omega_\qvec = cq$, we can switch from spectral densities in the
wavevector domain to spectral densities in the frequency domain.
In particular, we can consider the spectral energy density
\begin{equation}
 U(\omega) = \frac{\hbar\omega^3}{4\pi^3 c^3}\mathcal F
\end{equation}
which is the energy density of an infinite cavity in which each
mode is populated by one photon. The term $\mathcal F$ is a
constant expressing the limited field-of-view (FOV) of the device
and depending on how the blackbody radiation is coupled into the
detector in the specific experimental setup.

The absorption probability can then be evaluated in the frequency
domain as
\begin{eqnarray}
\label{Pabs}
 P^{opt}_{\alpha\alpha'} & = &
 \frac{2\pi}{\hbar^2}
 \int |g_{\alpha\alpha'}
 (\omega)|^2 n(\omega)\delta\left(\Delta\omega -
 \omega\right)\domega = \\ \nonumber
 & = & \frac{2\pi}{\hbar^2} |g_{\alpha\alpha'}(\Delta\omega)|^2 n(\Delta\omega)
\end{eqnarray}
where $\Delta\omega = \frac{E_\alpha - E_{\alpha'}}{\hbar}$ is the
resonance frequency of the transition.

After substitution of equations~(\ref{g3}) and~(\ref{A}) into
Eq.~(\ref{Pabs}), we obtain
\begin{equation}
 P^{opt}_{\alpha\alpha'} = U(\Delta\omega)\frac{\pi e^2}{m^2\varepsilon \Delta\omega^2}|p_{\alpha\alpha'}|^2 n(\Delta\omega) \delta(\kvec_\alpha -
 \kvec_{\alpha'}) \,.
\end{equation}
Since we are dealing with a thermal population of photons we take
$n(\Delta\omega)$ as the Bose-Einstein distribution function so
that the total absorption probability can be finally written as
\begin{equation}
\label{Popt}
 P^{opt}_{\alpha\alpha'} =\frac{ e^2\hbar\Delta\omega\mathcal F}{4\pi^2  c^3m^2\varepsilon}|p_{\alpha\alpha'}|^2 \frac{1}{e^{\frac{\hbar\Delta\omega}{k_BT}} - 1} \delta(\kvec_\alpha -
 \kvec_{\alpha'}) \; .
\end{equation}
Analogously, the emission probability is
\begin{equation}
\label{Popt2}
 P^{opt}_{\alpha'\alpha} =
 \frac{ e^2\hbar\Delta\omega\mathcal F}{4\pi^2  c^3m^2\varepsilon}|p_{\alpha'\alpha}|^2
 \left[\frac{1}{e^{\frac{\hbar\Delta\omega}{k_BT}} - 1} + 1\right]
 \delta(\kvec_{\alpha'} - \kvec_\alpha) \; .
\end{equation}

\subsection{Fixing the value of $\tau$}

The model contains a free parameter, $\tau$, which has to be
adjusted in order to reproduce some experimental data. Its value
is, in principle, crucial in determining the \Tblip\ of the
simulated devices, since changing the mean lifetime of electrons
will change the strength of thermal scattering with respect to
optical scattering and thus will affect the point at which these
two competing processes balance.

In particular, we choose to adjust $\tau$ in order to reproduce
the measured \Tblip\ (12 K) of the bound-to-continuum QWIP
operating at 3.2 THz and reported in Ref.~[\onlinecite{Liu2}].

\begin{figure}
  \includegraphics[width=\linewidth,height=.8\linewidth]{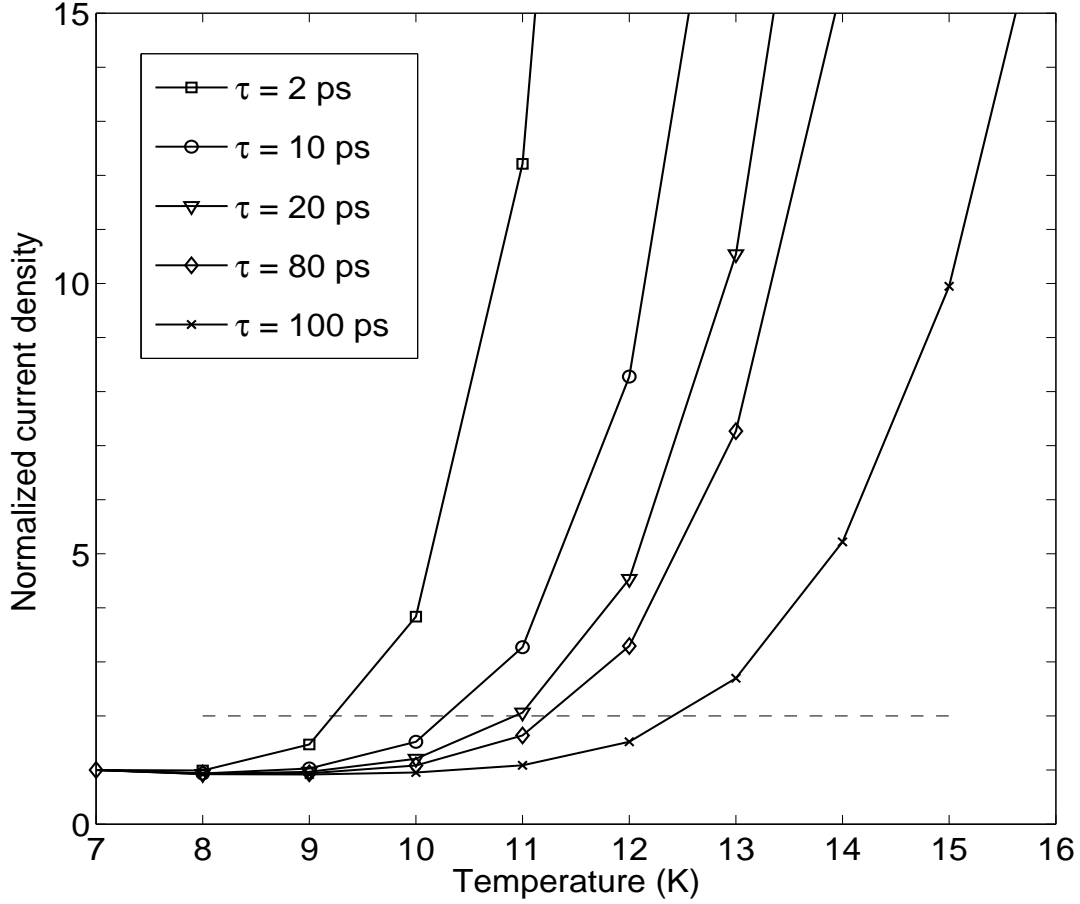}
  \caption{Estimated normalized current density along the growth direction, as a function of device temperature,
  for the one-level QWIP of Figure~1, operating at 3 THz in the presence of a background radiation field at 300 K.
  Different symbols correspond to different values of $\tau$. Applied electric field is 50 V/cm, FOV is 90$^\circ$.
  The dashed line marks the current doubling.}\label{tau_fit}
\end{figure}

Figure~\ref{tau_fit} shows the total normalized current densities
that we obtain for a 3 THz QWIP as a function of temperature for
different values of $\tau$. Although $\tau$ is the key parameter
that fixes the value of the \Tblip, it can be noted from the
figure that in the interval $\tau \approx 50-100$ ps the \Tblip\
shows little variation around 12 K. We can thus safely assume for
$\tau$ any value in this range, like, e.g., $\tau = 80$ ps, in
order to reproduce the experimental data.

It is important to stress once more, at this point, that this very
large value derives from the fact that we are using a simplified
model for thermal scattering; our fitting parameter $\tau$ is not
to be taken as a realistic indication of electron scattering time
in the real heterostructure.

Once the value of $\tau$ has been set, on the basis of the above
discussion, we use it in modeling the current response of
detectors operating at identical frequencies but employing the
proposed bound-to-bound-to-continuum strategy, and differing in
the number of bound states.

\section{Results and discussion}

We now apply the model of Section II to describe four different
devices, having a number of bound levels ranging from one
(standard QWIP) to four, and designed according to our
bound-to-bound-to-continuum strategy. All devices are exposed to a
300~K blackbody radiation under a 90$^\circ$ FOV and are subject
to a 50~V/cm external bias.


\begin{figure}
  \includegraphics*[width=\linewidth,height=.8\linewidth]{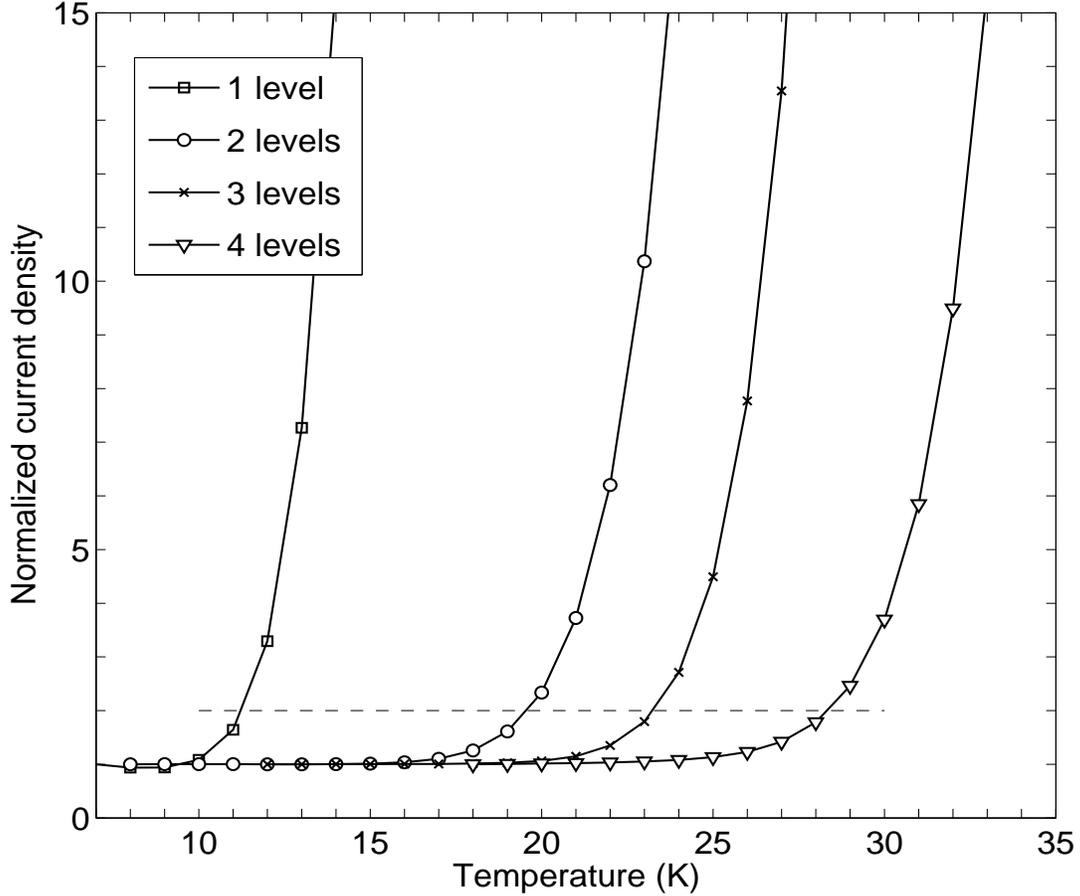}
  \caption{Estimated current density (normalized) along the growth direction,
  as a function of device temperature, for the four diverse multilevel designs of Figure~1,
  differing in the number of bound states.}\label{jzTN}
\end{figure}

Figure~\ref{jzTN} shows the total normalized currents across each of the four
devices as a function of the device temperature. Each curve allows
to identify a low-temperature regime in which the dark current is
negligible with respect to the photocurrent: the total current is
therefore independent from the device temperature. Conversely, in
the high temperature region, the dark current increases almost
exponentially so that the photocurrent quickly becomes negligible
and the current is totally due to the `dark' contribution. The
\Tblip\ may be identified as the temperature at which the total
current doubles with respect to the low temperature region (dashed horizontal line in Fig. \ref{jzTN}): at
this temperature the dark current and the photocurrent have the
same magnitude.


\begin{figure}
  \includegraphics*[width=\linewidth,height=.8\linewidth]{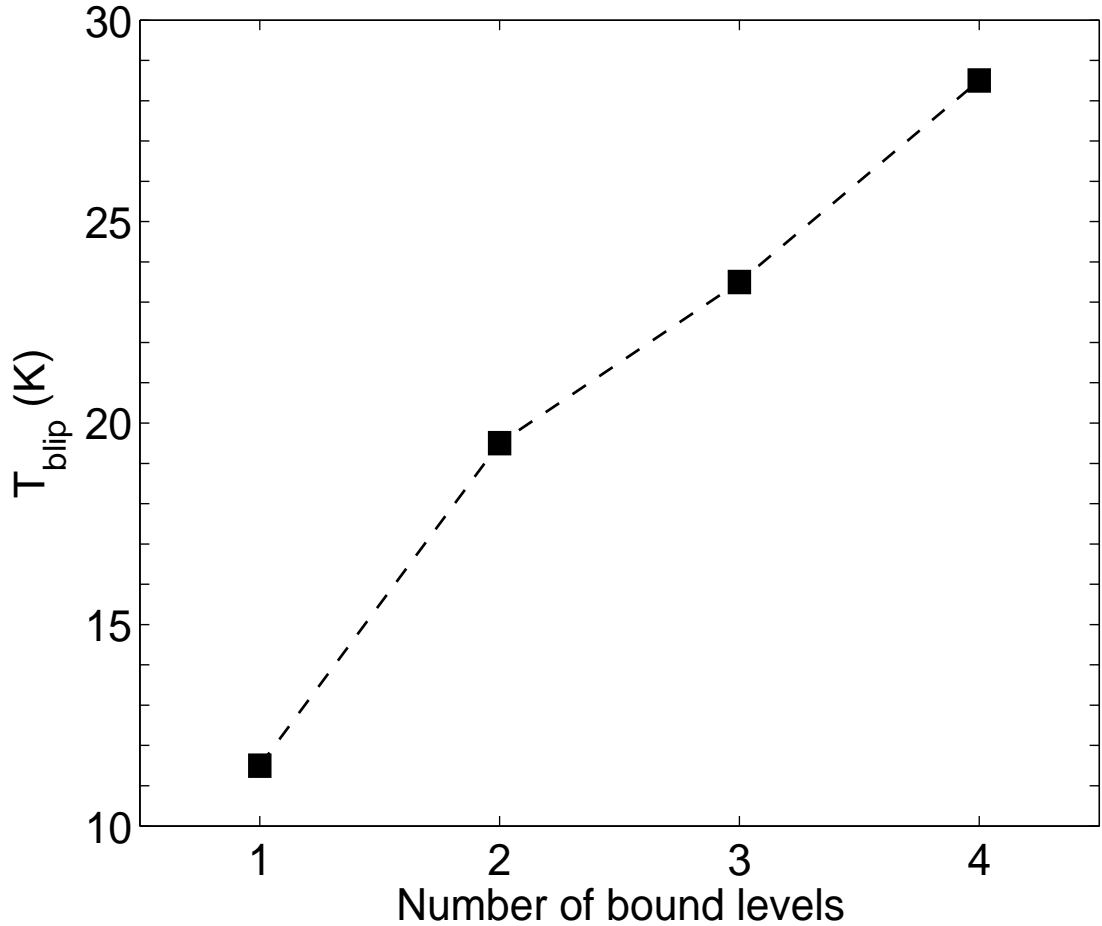}
  \caption{Estimated values of \Tblip\ for the four devices of Figure~1, as deduced from the
  data shown in Figure~\protect\ref{jzTN}. The dashed line is a guide to the eye.} \label{Tblip_vs_N}
\end{figure}

The diverse designs have values of \Tblip\ = 11.5, 19.5, 23.5 and
28.5 K for one, two, three and four bound levels, respectively,
showing the trend reported in Figure~\ref{Tblip_vs_N}.

\begin{figure}
  \includegraphics*[width=\linewidth,height=.8\linewidth]{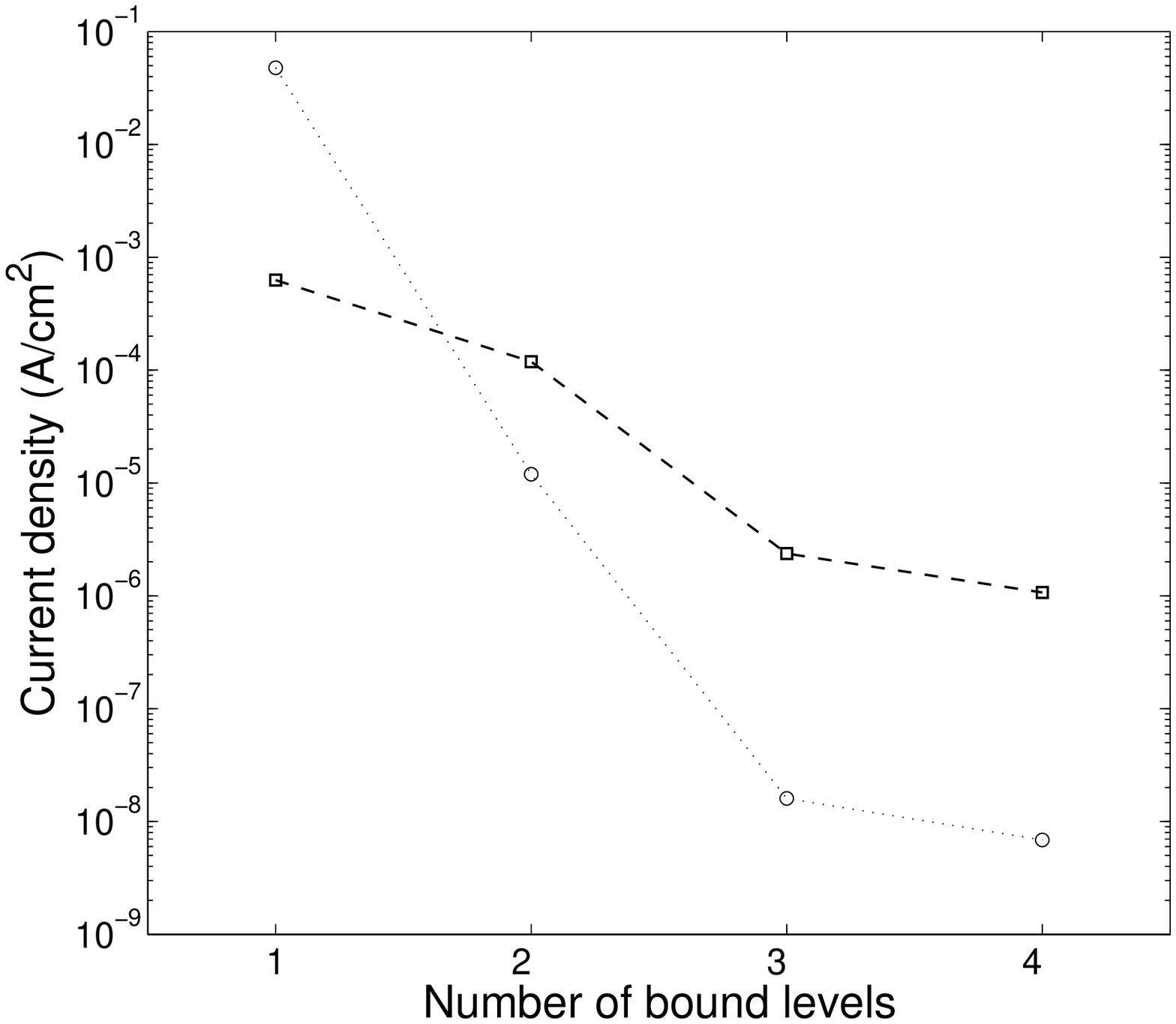}
  \caption{Photocurrent (squares) and dark current at 17 K (circles) as a function of the
  number of levels in the device. The dark current decreases more rapidly than the
  photocurrent, on increasing the number of bound subbands, and thus the \Tblip\ increases.
  Lines are a guide to the eye.} \label{jzjdN}
\end{figure}

The increase of \Tblip\ may be better interpreted by looking at
Figure~\ref{jzjdN}, where the photocurrent and the dark current
are plotted as a function of the number of bound levels. Both
currents decrease on increasing the latter, but the dark current
does it faster. Therefore, the temperature at which the two are
equal moves towards higher values.

As can be observed in Fig. \ref{jzjdN} there is a dramatic
decrease in the photocurrent when switching from two to three
bound states, which is mainly due to the reduction of the
photoconductive gain. In fact, for the four-level design, the
latter reduces to just the 0.2\% of the value of the one-level
QWIP. Conversely the quantum efficiency is only lowered by 14\%
and thus its variation does not significantly affect the
photocurrent.

This behavior can be explained by considering Fig. \ref{profile}
and noting that there is a remarkable geometrical difference
between the two-level and the three-level design. The presence of
the nested QW introduces a new ground state whose wavefunction has
little overlap with the wavefunctions of higher energy states,
therefore reducing the oscillator strength. This conclusion is
supported by the fact that the photocurrent reduction between the
three and four levels designs, where no major structural change
has been introduced, is comparable to the decrease between one and
two levels cases. In the present work the optimization of device
performance is not the central issue; in this respect, a more
elaborate tuning of the device geometry to achieve higher
oscillator strengths would surely allow for better operational
results.

\section{Summary and conclusions}
The scaling-down of QWIPs to access the terahertz range of the
electromagnetic spectrum is not straightforward: in this frequency
range the dark current, mainly due to the high-energy tail of the
electron distribution function, may indeed become predominant over
the photocurrent signal. In a recent paper~\cite{multi10} we have
proposed and theoretically investigated a THz-detector design
alternative to the conventional QWIP structure. The former,
instead of resorting on the conventional bound-to-continuum
scheme, exploits a bound-to-bound-to-continuum strategy. In
particular, a ladder of equally-spaced bound levels is employed,
whose energy-step is tuned to the desired detection frequency.

Our previous analysis demonstrated that a multilevel architecture
can indeed satisfactorily face the dark-current problem in
far-infrared QWIPs. In the present paper we have significantly
improved some features of our model to better reproduce the
behavior of realistic state-of-the art designs. In particular, our
attention has been devoted to a specific figure of merit of QWIPs,
such as the background-limited infrared photodetection temperature
($T_{\rm blip}$), which is related to the interplay between dark
current and photocurrent. Our results have demonstrated that the
proposed multi-subband scheme allows for higher \Tblip\ values,
with respect to conventional QWIP designs operating at the same
frequency, and therefore could represent a better alternative for
THz radiation detection.

\appendix

\section{Potential profile calculation}
\label{appA}
The key-point of the proposed bound-to-bound-to-continuum
architecture is the design of the nanostructure potential profile.
The latter requires the solution of the inverse problem of setting
the desired energy spectrum and then finding the corresponding
operator, i.e., the potential energy term of the electron
hamiltonian $H_e^\circ$. In the present paper, this problem has
been solved numerically, by means of a variational approach.

In particular, starting from the function $V(z)$ that describes
the potential profile in one period of our device, a functional
$\Fcal[V(z)]$ has been defined whose value represents how far the
function $V(z)$ is from our target function $\tilde V(z)$. The
latter must be such that the operator
\begin{equation}
 H = -\frac{\hbar^2}{2m}\DDfDxx{z} + \tilde V(z)
\end{equation}
has a spectrum composed of a lower part with $\tilde N$
equally-spaced discrete values (bound states) and an upper
continuous part. The number $\tilde N$ and the energy spacing
$\tilde E$ of the bound states are our design constraints.

In general, a function $V(z)$ will produce $N$ bound states of
energies $E_i$, $i = 1, \dots, N$, where $N$ can range from one to
infinity (infinitely deep potential well). If there are two or
more bound states we can define a mean interlevel spacing
\begin{equation}
 E = \frac{1}{N}\sum_{i=1}^N \left(E_{i+1} - E_i \right)
\end{equation}
and a level spreading
\begin{equation}
 \sigma = \frac{1}{N} \sum_{i=1}^N (E_i - E)^2 \; .
\end{equation}

In terms of the quantities $E$ and $\sigma$ specified above, the
functional $\Fcal[V(z)]$ is defined as
\begin{equation}
 \Fcal[V(z)] = (1 - \delta_{N\tilde N}) + \sigma + |E - \tilde E|
 \; .
\end{equation}
From the latter equation we see that $\Fcal$ is always positive
and assumes the minimum, null, value only when $N = \tilde N$,
$\sigma = 0$ and $E = \tilde E$.

The actual existence of the minimum depends on the functional
space we choose for $V(z)$. Indeed, we already know a solution for
the problem $\Fcal = 0$ which is the harmonic oscillator, but the
latter cannot be taken into consideration because it is not a
realistic potential profile and its spectrum does not contain a
continuous part.

Without going into the rigorous mathematical definition of the
space, which is beyond the scope of the present paper, $V(z)$ must
be a periodic function with period $L_z$. We therefore define
$V(z)$ on the domain $-L_z/2 < z < L_z/2$ and impose $V(-L_z/2) =
V(L_z/2)$. Of course we don't want $V(z)$ to diverge at any point
and in addition we want it to be as close as possible to realistic
and technologically accessible potential profiles.

We choose to take $V(z)$ piecewise constant on its domain so that
it can be described by a discrete set of $M$ parameters
representing widths and depths of every constant sector. In this
way, $V(z)$ can be represented by a point in an $M$-dimensional
space and $\Fcal$ actually becomes a function of $M$ variables. In
practice, $V(z)$ takes the form of a multi quantum well or a
nested quantum well structure, in which we vary depths and widths
of the diverse layers.

The minimization of $\Fcal$ is not trivial mainly because of the
$\delta_{N \tilde N}$ term that makes it discontinuous in an
unpredictable way: by slowly varying the free parameters, the
potential profile can suddenly produce a new bound state or lose
one causing $\Fcal$ to jump by $\pm 1$ and thus we cannot use
methods that seek a local minimum following the function gradient.

We adopted the easiest possible solution: starting from an initial
guess for $V(z)$, the free parameters are varied within a certain
range to see if a minimum is present and whether the latter is
actually the absolute one, for which $\Fcal = 0$. The existence of
such minima mostly depends on the number $M$ of free parameters
that can be varied. We choose to start with the minimum number of
parameters (which is two for a single quantum well) and then
gradually increase this number in order to generate more bound
states.


\section{Boltzmann equation solution}
\label{appB}
\subsection{State space discretization}
The electron dynamics in our prototypical quantum device is
described by the Boltzmann transport equation~(\ref{boltzmann2}).
The latter will be solved by finite difference discretization of
the derivatives and Reimann discretization of the integral.

Due to the cylindrical symmetry of the physical problem,
guaranteed, as in our case, by an external field applied along the
growth direction only, a convenient starting point is to employ
cylindrical coordinates, with $k_z$ being the perpendicular
(growth direction) wavevector, and $k_p$ and $\theta$ the modulus
and anomaly, respectively, of the in-plane wavevector.

A central difference approximation of the derivatives along $k_z$
may then be applied, with periodic boundary conditions accounting
for the repetition of the Brillouin zone. In particular, the
values of $k_z$ span the first Brillouin zone $-\frac{\pi}{L_z} <
k_z < \frac{\pi}{L_z}$ forming a uniform grid of step $N_{k_z}$.
The width of each discrete cell is therefore equal to $\Delta k_z
= \frac{2\pi}{L_z N_{k_z}}$.

The in-plane angle $\theta$ is uniformly discretized in the domain
$[0,2\pi)$. The number of discrete cells is $N_\theta$ and their
size is $\Delta \theta = \frac{2\pi}{N_\theta}$. Again, central
difference approximation of derivatives and periodic boundary
conditions are adopted.

Discretization along $k_p$ poses the problem of limiting the
in-plane $\kvec$-space. The electron distribution function
$f(k_p)$ at thermal equilibrium has the form
\begin{equation}
 f(k_p) \propto e^{- \frac{\hbar^2 k_p^2}{2m k_B T}}
\end{equation}
and thus decays rather quickly in $k_p$. We expect the electron
non-equilibrium distribution to decay more or less in the same way
in the presence of the external radiation field. We therefore set
a cutoff value, $f_{\textrm{cut}}$, below which $f(k_p)$ is
considered to be negligible, and use it to define a maximum value
for $k_p$ in the following way
\begin{equation}
 k_{p}^{\textrm{max}} = \sqrt{\frac{2mk_BT}{\hbar^2}\ln
 f_{\textrm{cut}}} \; .
\end{equation}
The interval $[0,k_{p}^{\textrm{max}}]$ is then discretized into a
uniform grid of dimension $N_{k_p}$, and derivatives are
approximated by central difference formulae in the inner nodes. In
particular, $f(k_p = 0)$ is supposed to have null derivative
(gaussian-like behavior) and the same applies for
$f(k_{p}^{\textrm{max}})$. Discrete cells along $k_p$ have a width
$\Delta k_p = \frac{k_{p}^{\textrm{max}}}{N_{k_p}}$.

To complete the description of the state-space we need to set the
number $N_b$ of subbands actually considered for calculations. The
plane-wave solution of the Schr\"odinger equation requires from
one to two hundred plane waves to give stable energy level values,
producing the same number of subbands. However, the electron
distribution function decays rather quickly and a number of bands
from five to a few tens is usually enough to ensure convergence.
Indeed, the actual number of bands depends on the operating
conditions of the device, such as the temperature and the presence
of incident light or external bias.

\subsection{Discrete Boltzmann equation}
After discretization of the state-space, the distribution function
$f_b(\kvec)$ can be itself discretized into a vector of components
$f_i$. The label $i$ ranges from $1$ to  the total number of grid
points $N = N_{k_z}N_{k_p}N_\theta N_b$ and accounts for the band
index $b_i$ and the three $\kvec$-space coordinate indexes
$k_{z,i}$, $k_{p,i}$ and $\theta_i$, collectively named $\kvec_i$.
The value $f_i$ is the mean of $f_{b_i}(\kvec_i)$ over the grid
volume element $\Delta\kvec_i = \Delta k_{z,i} \Delta k_{p,i}
\Delta \theta_i$
\begin{equation}
 f_i = \frac{1}{\Delta\kvec_i}\int_{\Delta\kvec_i}f_{b_i}(\kvec_i)\dkvec
\end{equation}
from $f_i$ we define the occupation number $n_i$ of the $i-$th discrete cell as
\begin{equation}
 n_i = f_i\Delta\kvec_i \quad .
\end{equation}
With this definition and taking into account relation (\ref{fnorm}) the occupation number is normalized as
\begin{equation}
\sum_in_i = N_e \; .
\end{equation}

After the discretization of the distribution function, we need to
find a suitable discretization of the scattering probabilities
$P_{bb'}(\kvec,\kvec')$. The total number of particles $R_{ji}$
that perform a transition from the volume $\Delta\kvec_i$ in band
$b_i$ to the volume $\Delta\kvec_j$ in band $b_j$, is given by the
probability that a particle in $\kvec_i$ performs a transition
towards one of the states in volume $\Delta\kvec_j$, which is
$\int_{\Delta\kvec_j}P_{b_jb_i}(\kvec,\kvec_i)\dkvec$, integrated
over all states of the starting volume
\begin{equation}
\label{R1}
 R_{ji} =
 \int_{\Delta\kvec_i}\int_{\Delta\kvec_j}f_{b_i}(\kvec')P_{b_jb_i}(\kvec,\kvec')\dkvec\dkvec'
 \, .
\end{equation}
In the discretized system the probability that a particle performs
the same transition is $P_{ji}\Delta\kvec_j$ and the number of
particles in the starting volume is $n_i = f_i\Delta\kvec_i$, thus
\begin{equation}
\label{R2} R_{ji} = f_i\Delta\kvec_iP_{ji}\Delta\kvec_j \, .
\end{equation}

The combination of equations~(\ref{R1}) and~(\ref{R2}) allow us to
derive the following expression for $P_{ji}$
\begin{equation}
\label{Pji}
 P_{ji} =
 \frac{1}{f_i\Delta\kvec_i\Delta\kvec_j}\int_{\Delta\kvec_i}\int_{\Delta\kvec_j}f_{b_i}(\kvec')P_{b_jb_i}(\kvec,\kvec')\dkvec\dkvec'
 \; .
\end{equation}
If we approximate $f_{b_i}(\kvec')$ with its mean value $f_i$ over
the volume $\Delta\kvec_i$, we can take it out of the integral and
rewrite Eq.~(\ref{Pji}) as
\begin{equation}
\label{Wji}
 W_{ji} = P_{ji}\Delta\kvec_j = \frac{1}{\Delta\kvec_i}\int_{\Delta\kvec_i}\int_{\Delta\kvec_j}P_{b_jb_i}(\kvec,\kvec')\dkvec\dkvec'
\end{equation}
and Eq.~(\ref{R2}) as
\begin{equation}
 R_{ji} = W_{ji}n_i \, .
\end{equation}
The quantity $W_{ji}$ is the probability that an event bringing an
electron from a state $i$ to one of the states in volume
$\Delta\kvec_j$ occurs. By multiplying $W_{ji}$ by the number of
particles in the volume $\Delta\kvec_i$, that is $n_i$, one
obtains the rate $R_{ji}$ of particles leaving the volume
$\Delta\kvec_i$ and entering the volume $\Delta\kvec_j$.

For the generic $i$-th volume element we can then write a rate
equation in the usual form
\begin{equation}
 \DfDx[n_i]{t} = \sum_j(W_{ij}n_j - W_{ji}n_i)
\end{equation}
which is the Boltzmann equation for a discrete system composed of
$N$ states. For the stationary state we write
\begin{equation}
\label{boltzmann3}
 \sum_j(W_{ij}n_j - W_{ji}n_i) = 0 \; .
\end{equation}

Due to the discretization procedure, the drift term in
Eq.~(\ref{boltzmann1}) can be written as
\begin{equation}
 \nabla f_b(\kvec)\cdot\frac{q}{\hbar}\Fvec = \sum_jW_{ij}^D f_j
\end{equation}
where $W_{ij}^D$ is an equivalent scattering matrix that can be
included into the $W_{ij}$ term in Eq.~(\ref{boltzmann3}). The
latter will in general consist of several contributions
\begin{equation}
 W_{ij} = W_{ij}^D + W_{ij}^{th} + W_{ij}^{opt}
\end{equation}
where $W_{ij}^{th}$ and $W_{ij}^{opt}$ are the discretized
probability densities $P^{th}_{\alpha\alpha'}$ and
$P^{opt}_{\alpha\alpha'}$ defined in Eqn.s~(\ref{Pth}) and
(\ref{Popt}) and computed using Eq.~(\ref{Wji}).

\subsection{Non-optical scattering probabilities}
\label{appC}
The derivation of Section \ref{Nopt} can be followed in the
discretized system by replacing $f_b(\kvec)$ with $f_i$ and
$P_{bb'}(\kvec,\kvec')$ with $P_{ij}$. In particular, the
discretized version of Eq.~(\ref{tau2})
\begin{equation}
\frac{1}{\tau} = \frac{1}{(2\pi)^3N_e}\sum_{ij} P^{th}_{ji}\Delta \kvec_j \Delta \kvec_i
\end{equation}
leads to a definition of $P^{th}_{ij}$ similar to Eq.~(\ref{Pth})
\begin{equation}
 P^{th}_{ij} = P_0\Pcal_{ij} = P_0
\begin{cases}
 1 \qquad \qquad \qquad \text{ if } E_j > E_i \\
  e^{-\frac{E_i - E_j}{k_BT}} \quad \text{ if } E_j < E_i
\end{cases}
\end{equation}
where again $P_0$ is a normalization constant that can be computed as
\begin{equation}
 \frac{1}{P_0} = \frac{\tau}{(2\pi)^3N_e}\sum_{ij} \Pcal_{ij} \Delta \kvec_j \Delta
 \kvec_i \, .
\end{equation}
The discrete thermal transition probabilities are then written
according to Eq~(\ref{Wji})
\begin{equation}
 W_{ij}^{th} = P_{ij}^{th}\Delta\kvec_i \, .
\end{equation}

\subsection{Optical scattering probabilities}
\label{appD}
The discrete optical scattering probabilities can be directly
computed using Eqns.~(\ref{Wji}) and~(\ref{Popt})
or~(\ref{Popt2}). In both cases, the transition probabilities are
of the form
\begin{equation}
 P^{opt}_{ij}(\kvec_i,\kvec_j) = w_{ij}\delta(\kvec_i - \kvec_j)
\end{equation}
where $w_{ij}$ contains all the coefficients and differs for
absorption and emission processes. Substituting the latter
expression into Eq.~(\ref{Wji}) gives
\begin{eqnarray}
 W^{opt}_{ij} & = & \frac{1}{\Delta\kvec_j}\int_{\Delta\kvec_j}\int_{\Delta\kvec_i} w_{ij}
 \delta(\kvec - \kvec')\dkvec'\dkvec \\ \nonumber
 & = &
 \frac{\delta_{ij}}{\Delta\kvec_j}\int_{\Delta\kvec_j}w_{ij}\dkvec
 \, .
\end{eqnarray}

Assuming that $w_{ij}$ is a smooth function over the volume cell
$\Delta\kvec_j$ and for a sufficiently dense grid, we can
approximate $w_{ij}$ as a constant and take it out of the
integral, which in turn results to be $\Delta\kvec_j$, therefore
yielding
\begin{equation}
 W^{opt}_{ij} = w_{ij}\delta_{ij} \; ,
\end{equation}
which is the discrete transition probability to be used in the
solution of the discrete Boltzmann equation.

\end{document}